\documentclass{www2006-submission}
\usepackage{graphicx}

\def\more-auths{%
\end{tabular}
\begin{tabular}{c}}
\def\sharedaffiliation{%
\end{tabular}
\begin{tabular}{c}}

\begin{document}
\title{How to make the top ten: \\ Approximating PageRank from in-degree}

\numberofauthors{2}

\author{
\alignauthor Santo Fortunato$^{1,2}$\\
     \email{santo@indiana.edu}
\alignauthor Mari\'{a}n Bogu\~{n}\'{a}$^{3}$\\
      \email{marian.boguna@ub.edu}
\vskip1pc  
\more-auths
\alignauthor Alessandro Flammini$^{1}$ \\
     \email{aflammin@indiana.edu}
\alignauthor Filippo Menczer$^{1}$ \\
     \email{fil@indiana.edu}
\vskip1pc  
\sharedaffiliation
\affaddr{$^{1}$ School of Informatics}  \\
\affaddr{Indiana University}   \\
\affaddr{Bloomington, IN 47406, USA}
\sharedaffiliation
\affaddr{$^{2}$ Fakult\"at f\"ur Physik}  \\
\affaddr{Universit\"at Bielefeld}   \\
\affaddr{D-33501 Bielefeld, Germany}
\sharedaffiliation
\affaddr{$^{3}$ Departament de F\'{i}sica Fonamental} \\
\affaddr{Universitat de Barcelona}  \\
\affaddr{08028 Barcelona, Spain}
}

\date{}

\maketitle
\begin{abstract}
PageRank has become a key element in the success of  
search engines, allowing to rank the most important hits 
in the top screen of results. One key aspect that distinguishes
PageRank from other prestige measures such as in-degree 
is its global nature. From the information provider 
perspective, this makes it difficult or impossible
to predict how their pages will be ranked. Consequently
a market has emerged for the optimization of search engine 
results. Here we study the accuracy with which PageRank 
can be approximated by in-degree, a local measure made 
freely available by search engines.  Theoretical and empirical 
analyses lead to conclude that given the weak degree 
correlations in the Web link graph, the approximation 
can be relatively accurate, giving service and information 
providers an effective new marketing tool. 
\end{abstract}




\section{Introduction}

PageRank has become a key element in the success of Web 
search engines, allowing to rank the most important hits 
in the top page of results. Certainly the introduction of 
PageRank as a factor in sorting results~\cite{Brin98} 
has contributed considerably to Google's lasting dominance 
in the search engine market~\cite{NielsenAug05}.

But PageRank is not the only possible measure of importance 
or prestige among Web pages. The simplest possible way to measure 
the prestige of a page is to count the incoming links (in-links) 
to the page. The number of in-links (in-degree) is the number 
of citations that a page receives from the other pages, so 
there is a correlation between in-degree and quality, especially 
when the in-degree is large. The in-degree of Web pages is very 
cheap to compute and maintain, so that a search engine can easily 
keep in-degree updated with the evolution of the Web. 

However, in-degree is a local measure.  All links to a page are
considered equal, regardless of where they come from.  Two pages with
the same in-degree are considered equally important, even if one is
cited by much more prestigious sources than the other.  To modulate the
prestige of a page with that of the pages pointing to it means to move
from the examination of an individual node in the link graph to that of
the node together with its predecessor neighbors.  PageRank represents
such a shift from the local measure given by in-degree toward a global
measure where each Web page contributes to define the importance of
every other page.

From the information provider perspective, the global nature of
PageRank makes it difficult or impossible to predict how a new page
will be ranked.  Yet it is vital for many service and information
providers to have good rankings by major search engines for relevant
keywords, given that search engines are the primary way that Internet
users find and visit Web sites~\cite{websidestory,Cho05WebDB}.  This
situation makes PageRank a valuable good controlled by a few major
search engines.  Consequently a demand has emerged for companies who
perform so-called \emph{search engine optimization} or \emph{search
engine marketing} on behalf of business clients.  The goal is to
increase the rankings of their pages, thus directing traffic to their
sites~\cite{sew-seo}.  Search engine marketers have partial knowledge
of how search engines rank pages.  They have access to undocumented
tools to measure PageRank, such as the Google toolbar.  Through
experience and empirical tests they can reverse-engineer some important
ranking factors.  However from inspecting the hundreds of bulletin
boards and blogs maintained by search engine marketers it is evident
that their work is largely guided by guesswork, trial and error.
Nevertheless search engine optimization has grown to be a healthy
industry as illustrated by a recent study~\cite{iprospect}.  Search
engine marketing has even assumed ethical and legal ramifications as a
sort of arms race has ensued between marketers who want to increase
their clients' rankings and search engines who want to maintain the
integrity of their systems.  The term \emph{search engine spam} refers
to those means of promoting Web sites that search engines deem
unethical and worthy of blocking~\cite{Gori05cacm}.

The status quo described above relies on two assumptions: (i) PageRank
is a quantitatively different and better prestige measure compared to
in-degree; and (ii) PageRank cannot be easily guessed or approximated
by in-degree.  There seems to be plenty of anecdotal and indirect
evidence in support of these assumptions --- for example the popularity
of PageRank --- but little quantitative data to validate them.  To wit,
Amento {\it et al.}~\cite{Amento00} report a very high average
correlation between in-degree and PageRank (Spearman $\rho=0.93$,
Kendall $\tau=0.83$) based on five queries.  Further, they report the
same average precision at 10 (60\%) based on relevance assessments by
human subjects.  In this paper we further quantitatively explore these
assumptions answering the following questions: \emph{What is the
correlation between in-degree and PageRank across representative
samples of the Web?  How accurately can one approximate PageRank from
local knowledge of in-degree?}

From the definition of PageRank, other things being equal, the PageRank
of a page grows with
the in-degree of the page.  Beyond this zero-order approximation, the
actual relation between PageRank and in-degree has not been thoroughly
investigated in the past.  It is known that the distributions of
PageRank and in-degree follow an almost identical
pattern~\cite{Upfal2002,millozzi}, i.e., a curve ending with a broad
tail that follows a power law with exponent about $2.1$.  This fact may
indicate a strong correlation between the two variables.  Surprisingly
there is no agreement in prior literature about the correlation between
PageRank and in-degree.  Pandurangan {\it et al.}~\cite{Upfal2002} show
very little correlation based on analysis of the Brown domain and the
TREC WT10g collection.  Donato {\it et al.}~\cite{millozzi} report on a
correlation coefficient which is basically zero based on analysis of a
much larger sample ($2\cdot 10^8$ pages) taken from the
WebBase~\cite{WebBase} collaboration.  On the other hand, analysis of
the University of Notre Dame domain by Nakamura~\cite{naka} reveals a
strong correlation.

In Section~\ref{meanfield} we estimate PageRank for a generic directed
network within a mean field approach.  We obtain a system of
self-consistent relations for the average value of PageRank of all
vertices with equal in-degree.  For a network without degree-degree
correlations the average PageRank turns out to be simply proportional
to the in-degree, modulo an additive constant.

The prediction is validated empirically in Section~\ref{results}, where
we solve the equations numerically for four large samples of the Web
graph; in each case the agreement between our theoretical estimate and
the empirical data is excellent.  We find that the Web graph is
basically uncorrelated, so the average PageRank for each degree class
can be well approximated by a linear function of the in-degree.  As an
additional contribution we settle the issue of the correlation between
PageRank and in-degree; the linear correlation coefficient is
consistently large for all four samples we have examined, in agreement
with Nakamura~\cite{naka}.  We also calculate the size of the
fluctuations of PageRank about the average value and find that the
relative fluctuations decrease as the in-degree increases, which means
that our mean field estimate becomes more accurate for important pages.

Our results suggest that we can approximate PageRank from in-degree. 
By deriving PageRank with our formula we can predict the rank of a page
within a hit list by knowing its in-degree and the number of hits in
the list.  Section~\ref{apps} reports on an empirical study of this
prediction, performed by submitting AltaVista~\cite{AltaVista} queries
to the Google API~\cite{GoogleAPI}.  The actual ranks turn out to be
scattered about the corresponding predictions.  The implication of the
fact that PageRank is mostly determined by in-degree is that it is
possible to estimate the number of in-links that a new page needs in
order to achieve a certain rank among all pages which deal with a
specific topic.  This provides search marketers --- and information
providers --- with a new powerful tool to guide their campaigns.

\section{Theoretical analysis}
\label{meanfield}

The PageRank $p(i)$ of a page $i$ is defined through the following
expression:
\begin{equation}
p(i)=\frac{q}{N}+(1-q)\sum_{j: j \rightarrow i}p(j)/k_{out}(j) \hskip0.7cm i=1,2, \dots, N
\label{eq1}
\end{equation}
where $N$ is the total number of pages, $j \rightarrow i$ indicates a
hyperlink from $j$ to $i$, $k_{out}(j)$ is the out-degree of page $j$
and $1-q$ is the so-called damping factor.  The set of
Equations~\ref{eq1} can be solved iteratively.
%
%
From Eq.~\ref{eq1} it is clear that the PageRank of a page grows with
the PageRank of the pages that point to it.  However, the sum over
predecessor neighbors implies that PageRank also increases with the
in-degree of the page.

PageRank can be thought of as the stationary probability of 
a random walk process with additional random jumps.  
The physical description of the
process is as follows: when a random walker is in a node of the
network, at the next time step with probability $q$ it jumps to a
randomly chosen node and with probability $1-q$ it moves to one of its
successors with uniform probability.  In the case of directed
networks, there is the possibility that the node has no successors. 
In this case the walker jumps to a randomly chosen node of the network
with probability one.  The PageRank of a node $i$, $p(i)$, is then the
probability to find the walker at node $i$ when the process has reached
the steady state, a condition that is always guaranteed by the
jumping probability $q$.

The probability to find the walker at node $i$ at time step $n$ follows a
simple Markovian equation:
\begin{eqnarray}
p_n(i) &=& \displaystyle{\frac{q}{N}+(1-q) \sum_{j:k_{out}(j)\neq 0}
\frac{a_{ji}}{k_{out}(j)}p_{n-1}(j)} \nonumber \\
&+& \displaystyle{\frac{1-q}{N} \sum_{j:k_{out}(j)= 0} p_{n-1}(j)},
\label{pagerank0}
\end{eqnarray}
where $a_{ji}$ is the adjacency matrix with entry $1$ if there is a
direct connection between $j$ and $i$ and zero otherwise.  The first
term in Eq.~\ref{pagerank0} is the contribution of walkers that decide
to jump to a randomly chosen node, the second term is the random walk
contribution, and the third term accounts for walkers that at the
previous step were located in dangling points and now jump to random
nodes.  In the limit $n\rightarrow \infty$ this last contribution
becomes a constant term affecting all the nodes in the same way and
thus, it can be removed from Eq.~\ref{pagerank0} under the constraint
that the final solution is properly normalized.  Hereafter we will use
this approach.  The PageRank of page $i$ is the steady state solution
of Eq.~\ref{pagerank0}, $p(i) = \lim_{n \rightarrow \infty} p_n(i)$.

Equation~\ref{pagerank0} can be used as a numerical algorithm to
compute PageRank but, unfortunately, it is not possible to extract
analytical solutions from it. In the next subsection we propose 
a mean field solution of Eq.~\ref{pagerank0} that, nevertheless, 
gives a very accurate description of the PageRank structure of the Web.

\subsection{Mean field analysis}

Instead of analyzing the PageRank of single pages, we aggregate pages
in classes according to their degree ${\bf k}
\equiv(k_{in},k_{out})$ and define the average PageRank of nodes of
degree class ${\bf k}$ as 
\begin{equation}
\overline{p}_n({\bf k}) \equiv \frac{1}{N P({\bf k})} \sum_{i\in
{\bf k}} p_n(i).
\label{aveq}
\end{equation}
Note that now ``degree class ${\bf k}$'' means all the nodes with
in-degree $k_{in}$ and out-degree $k_{out}$. Taking the average of
Eq.~\ref{pagerank0} for all nodes of the degree class ${\bf k}$ we
obtain 
\begin{eqnarray}
\lefteqn{\frac{1}{N P({\bf k})} \sum_{i\in
{\bf k}} p_n(i) = \frac{q}{N}} \nonumber \\
&+& \frac{(1-q)}{N P({\bf k})}
\sum_{i\in{\bf k}}\sum_{j:k_{out}(j)\neq 0}\frac{a_{ji}}{k_{out}(j)}p_{n-1}(j). 
\label{pagerank1a}
\end{eqnarray}
From Eq.~\ref{aveq} we see that the left-hand side of Eq.~\ref{pagerank1a}
is $\overline{p}_n({\bf k})$. In the right-hand side we split the sum 
over $j$ into  two sums, one over all the degree classes ${\bf k'}$ 
and the other over all the nodes within each degree class ${\bf k'}$. We get
\begin{equation}
\overline{p}_n({\bf k})=\frac{q}{N}+\frac{(1-q)}{N P({\bf k}) }
\sum_{{\bf k'}}\frac{1}{k'_{out}}\sum_{i\in {\bf k}} \sum_{j\in {\bf
k'}} a_{ji}p_{n-1}(j). \label{pagerank1}
\end{equation}
At this point we perform our mean field approximation, 
which consists in substituting the PageRank of
the predecessor neighbors of node $i$ by its mean value, that is,
\begin{eqnarray}
\displaystyle{\sum_{i\in {\bf k}} \sum_{j\in {\bf k'}}
a_{ji}p_{n-1}(j)} &\simeq& \displaystyle{\overline{p}_{n-1}({\bf k'})
\sum_{i\in {\bf k}} \sum_{j\in {\bf k'}}
a_{ji}} \nonumber \\
 &=&\displaystyle{\overline{p}_{n-1}({\bf k'}) E_{{\bf k'}\rightarrow{\bf
 k}},}
\label{meanfiel}
\end{eqnarray}
where $E_{{\bf k'}\rightarrow{\bf k}}$ is the total number of
links pointing from nodes of degree ${\bf k'}$ to nodes of
degree ${\bf k}$. This matrix can also be rewritten as
\begin{eqnarray}
E_{{\bf k'}\rightarrow{\bf k}} &=& k_{in} P({\bf k})N \frac{E_{{\bf
k'}\rightarrow{\bf k}}}{k_{in} P({\bf k})N} \nonumber \\
&=& k_{in} P({\bf k})N
P_{in}({\bf k'}|{\bf k}), 
\label{E_k'k}
\end{eqnarray}
where $P_{in}({\bf k'}|{\bf k})$ is the probability that a 
predecessor of a node belonging to 
degree class ${\bf k}$ belongs to degree class ${\bf
k'}$. Using Equations~\ref{meanfiel} and \ref{E_k'k} in
Eq.~\ref{pagerank1} we finally obtain
\begin{equation}
\overline{p}_n({\bf k})=\frac{q}{N}+(1-q) k_{in} \sum_{{\bf
k'}}\frac{P_{in}({\bf k'}|{\bf k})}{k'_{out}}
\overline{p}_{n-1}({\bf k'}), \label{pagerank2}
\end{equation}
which is a closed set of equations for the average PageRank of pages in the
same degree class. When the network has degree-degree correlations,
the solution of this equation is non-trivial and the resulting
PageRank can have a complex dependence on the degree. However,
in the particular case of uncorrelated networks 
the transition probability
$P_{in}({\bf k'}|{\bf k})$ becomes independent of the degree ${\bf
k}$ and takes the simpler form
\begin{equation}
P_{in}({\bf k'}|{\bf k})=\frac{k'_{out} P({\bf k'})}{\langle k_{in}
\rangle}.
\end{equation}
Using this expression in Eq.~(\ref{pagerank2}) and taking the limit
$n\rightarrow \infty$, we obtain
\begin{equation}
\overline{p}({\bf k})=\frac{q}{N}+\frac{1-q}{N}
\frac{k_{in}}{\langle k_{in} \rangle},
\label{cleq}
\end{equation}
that is, the average PageRank of nodes of degree class ${\bf k}$ is
independent of $k_{out}$ and proportional to $k_{in}$.

\subsection{Fluctuation analysis}
\label{fluct}

The formalism presented in the previous subsection gives a solution for
the average PageRank of nodes of the same degree class but it tells us
nothing about how PageRank is distributed within one degree class.  To
fill this gap, we extend our mean field approach to the fluctuations
within a degree class.  To this end, we first start by taking the
square of Eq.~\ref{pagerank0}:
\begin{eqnarray}
  \lefteqn{p^2_n(i)  =  \displaystyle{\frac{q^2}{N^2}+\frac{2q(1-q)}{N} \sum_j
  \frac{a_{ji}}{k_{out}(j)}p_{n-1}(j)}} \nonumber \\
  &+& \displaystyle{
    (1-q)^2 \sum_j \frac{a_{ji}}{k_{out}^2(j)}p^2_{n-1}(j)} \nonumber \\
  &+& \displaystyle{
    (1-q)^2 \sum_{j\neq j'} \frac{a_{ji}
    a_{j'i}}{k_{out}(j)k_{out}(j')}p_{n-1}(j) p_{n-1}(j')}.
\label{pagerank02}
\end{eqnarray}
As in the previous calculation, we take the average over degree
classes of the square of PageRank and define
\begin{equation}
\overline{p^2}_n({\bf k})\equiv \frac{1}{N P({\bf k})} \sum_{i\in
{\bf k}} p^2_n(i).
\end{equation}
Taking this average in Eq.~\ref{pagerank02} and rearranging terms we get
\begin{eqnarray}
\lefteqn{\displaystyle{
 \overline{p^2}_n({\bf
k}) = \frac{q^2}{N^2}+\frac{2q(1-q)}{N} k_{in}\sum_{{\bf
k'}}\frac{P_{in}({\bf k'}|{\bf k})}{k'_{out}}
\overline{p}_{n-1}({\bf k'})}} \nonumber \\
 &+& \displaystyle{
 (1-q)^2 k_{in}\sum_{{\bf k'}}\frac{P_{in}({\bf
k'}|{\bf
k})}{k'^2_{out}} \overline{p^2}_{n-1}({\bf k'})} \nonumber \\
 &+& (1-q)^2 k_{in}(k_{in}-1) \cdot \nonumber \\
 & & \cdot \displaystyle{\sum_{{\bf k'}} \sum_{{\bf
k''}}\frac{P_{in}({\bf k'},{\bf k''}|{\bf k})}{k'_{out}k''_{out}}
\overline{p}_{n-1}({\bf k'}) \overline{p}_{n-1}({\bf k''})}, 
\end{eqnarray}
where we have used again the mean field approach. The probability
$P_{in}({\bf k'},{\bf k''}|{\bf k})$ is the joint probability that a
node of degree ${\bf k}$ has simultaneously one predecessor of
degree ${\bf k'}$ and another of degree ${\bf k''}$. We can make the 
further assumption that this joint distribution
factorizes as $P_{in}({\bf k'},{\bf k''}|{\bf k})=P_{in}({\bf
k'}|{\bf k})P_{in}({\bf k''}|{\bf k})$. In this case we can write
an equation for the standard deviation within a degree class,
$\sigma_n^2({\bf k})=\overline{p^2}_{n}({\bf
k})-\overline{p}_{n}^2({\bf k})$, as follows:
\begin{eqnarray}
\frac{\sigma_n^2({\bf k})}{(1-q)^2} &=& \displaystyle{k_{in}\sum_{{\bf
k'}}\frac{P_{in}({\bf k'}|{\bf k})}{k'^2_{out}} \sigma_{n-1}^2({\bf
k'})} \nonumber \\
&+& \displaystyle{k_{in} \sum_{{\bf k'}}\frac{P_{in}({\bf k'}|{\bf
k})}{k'^2_{out}} \overline{p}_{n-1}^2({\bf k'})} \nonumber \\
&-& \displaystyle{k_{in}\left[
\sum_{{\bf k'}}\frac{P_{in}({\bf k'}|{\bf k})}{k'_{out}}
\overline{p}_{n-1}({\bf k'})\right]^2}.
\label{compfleq}
\end{eqnarray}

In the case of uncorrelated networks, this equation can be
analytically solved in the limit $n\rightarrow \infty$:
\begin{equation}
\sigma^2({\bf k})= \frac{(1-q)^2}{N^2 \langle k_{in}
\rangle^2} \displaystyle{\frac{\frac{1}{\langle k_{in} \rangle}
\left\langle \frac{(q \langle k_{in}
\rangle + (1-q)k_{in})^2}{k_{out}}\right\rangle -1}{1-\frac{(1-q)^2}{\langle
k_{in}\rangle} \left\langle \frac{k_{in}}{k_{out}} \right\rangle} k_{in} }.
\label{simpfl}
\end{equation}
In the case of the Web, the heavy tail of the in-degree distribution
and the high average in-degree allows to simplify this expression as
\begin{equation}
\sigma^2({\bf k})\simeq \frac{(1-q)^4}{N^2 \langle k_{in}\rangle^3}
\left\langle \frac{k_{in}^2}{k_{out}}\right\rangle k_{in}.
\end{equation}
For large in-degrees, the coefficient of variation is
\begin{equation}
\frac{\sigma({\bf k})}{\overline{p}({\bf k})} \simeq (1-q) \left[
\left\langle \frac{k_{in}^2}{k_{out}}\right\rangle \frac{1}{\langle k_{in}
\rangle k_{in}} \right]^{1/2}.
\label{fleq}
\end{equation}
The factor $\left\langle \frac{k_{in}^2}{k_{out}}\right\rangle$ in this
expression can be very large if the network is scale-free, which
implies that the relative fluctuations are large for small
in-degrees. However, for large in-degrees the relative fluctuations
become less important --- due to the factor $k_{in}$ in the
denominator --- and the average PageRank obtained in the previous
subsection gives a good approximation. This can be seen by
analyzing the coefficient of variation for the nodes with the
maximum degree $k_{in}^{max}$. Assuming that $k_{out}$ is weakly
correlated with $k_{in}$, the coefficient $\left\langle
\frac{k_{in}^2}{k_{out}}\right\rangle$ scales with the maximum in-degree
as $(k_{in}^{max})^{3-\gamma_{in}}$ and the coefficient of
variation as $(k_{in}^{max})^{1-\gamma_{in}/2}$. Since
$\gamma_{in}>2$, the relative fluctuations go to zero. Then, for
small in-degrees we expect PageRank to be distributed according to a
power law; for intermediate in-degrees, according to a distribution
peaked at the average mean field value plus a power law tail; and for
large in-degrees, according to a Gaussian distribution centered
around the predicted mean field solution.

\section{Results}
\label{results}

We analyzed four samples of the Web graph. Two of them 
were obtained by crawls performed in 2001 and 2003
by the WebBase collaboration~\cite{WebBase}. The other two
were collected by the WebGraph project~\cite{WebGraph} using 
the UbiCrawler~\cite{ubicrawler}: the pages belong to two national domains, 
\texttt{.uk} (2002) and \texttt{.it} (2004), respectively. In Table~\ref{tab1} we 
list the total number of vertices and edges and the average degree for 
each data set.

\begin{table}[t]
\centering
\caption{\label{tab1} Number of pages, links, and average degree ($\langle k \rangle = \langle k_{in} \rangle = \langle k_{out} \rangle$) for the four data sets we have analyzed.
}
\vskip0.2cm
\begin{tabular}{|c|cccc|} \hline
Data set & WB 2001 & \texttt{.uk} 2002 & WB 2003 & \texttt{.it} 2004\\ \hline
\# pages & $8.1 \times 10^{7}$ & $1.9 \times 10^{7}$ & $4.9 \times 10^{7}$ & $4.1 \times 10^{7}$ \\ 
\# links & $7.5 \times 10^{8}$ & $2.9 \times 10^{8}$ & $1.2 \times 10^{9}$ & $1.1 \times 10^{9}$ \\ 
$\langle k \rangle$ & 9.34 & 15.78 & 24.05 & 27.50 \\ \hline
\end{tabular}
\end{table}

\begin{figure}[b]
\centering
\includegraphics[width=\columnwidth]{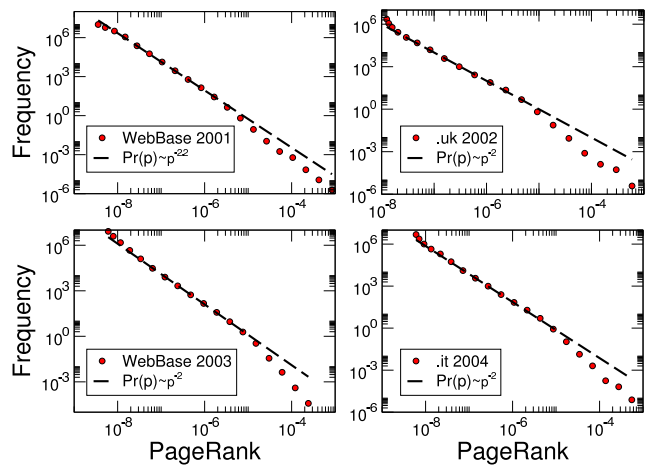} 
\caption{\label{fig1}PageRank distributions.}
\end{figure}

We calculated PageRank with the standard iterative procedure; the factor $q$ was 
set to $0.15$, as in the original paper by Brin and Page~\cite{Brin98} 
and many successive studies.
The convergence of the algorithm is very quick: in each case 
less than a hundred iterations were enough to determine the result with
a relative accuracy of $10^{-5}$ for each vertex.
In Fig.~\ref{fig1} we show the distributions of PageRank. In all four cases we 
obtained a pattern with a broad tail. The initial part of the distribution 
can be well fitted by a power law $p^{-\beta}$ with exponent $\beta$ between $2.0$ and $2.2$. 
This is in agreement with the findings of refs.~\cite{Upfal2002,millozzi}.
The right-most part of each curve,   
corresponding to the pages with highest PageRank, decreases faster.   
For the WebBase sample of $2001$  
the tail of the curve up to the last point can be well fitted by a power law 
with exponent $\beta \approx 2.6$; in the other cases we see evidence of an exponential 
cutoff. 

\begin{table}[t]
\centering
\caption{\label{tab2} Exponents of the power law part of the PageRank distribution and 
linear correlation coefficients between PageRank and in-degree.}
\vskip0.2cm
\begin{tabular}{|c|cccc|} \hline
Data set & WB 2001 & \texttt{.uk} 2002 & WB 2003 & \texttt{.it} 2004\\ \hline
$\beta$ & $2.2 \pm 0.1$ & $2.0 \pm 0.1$ & $2.0 \pm 0.1$ & $2.0 \pm 0.1$ \\ 
$\rho$ & 0.538 & 0.554 & 0.483 & 0.733 \\ \hline
\end{tabular}
\end{table}

We have also calculated the linear correlation coefficient between PageRank and in-degree.
In Table~\ref{tab2} we list Pearson's $\rho$ together with the slope of the power law
portions of the PageRank distributions. We see that the correlation between PageRank and in-degree 
is rather strong, in contrast to the findings of refs.~\cite{Upfal2002} and especially 
\cite{millozzi} but in agreement with ref.~\cite{naka}.

\begin{figure}[t]
\centering
\includegraphics[width=\columnwidth]{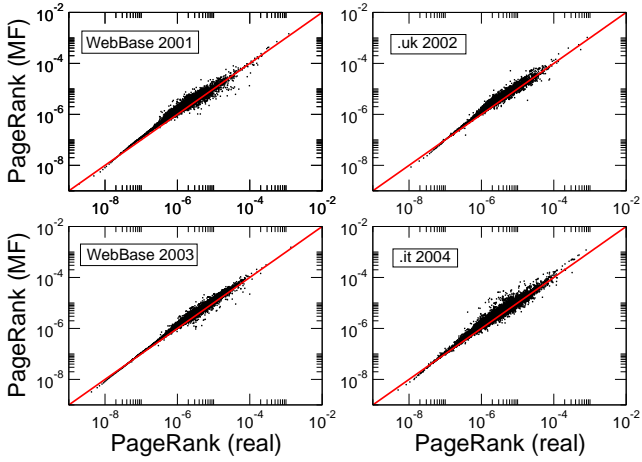}
\caption{\label{fig2} Scatter plots of the empirical average PageRank per degree
class versus our mean field (MF) estimate.}
\end{figure}

We solved Eq.~\ref{pagerank2} with an analogous iterative procedure 
as the one we used to calculate PageRank. We now look for the vector 
$\bar{p}({\bf k})$, defined for all
pairs ${\bf{k}}\equiv(k_{in},k_{out})$ which occur in the network. 
Since PageRank is a probability, it must be normalized so that its sum 
over all vertices of the network is one. So we initialized the vector with the 
constant $\bar{p}_0({\bf k})=1/N$, and plugged it into the right-hand 
side of Eq.~\ref{pagerank2} to get the first approximation $\bar{p}_1({\bf k})$. 
We then used $\bar{p}_1({\bf k})$ as input to get $\bar{p}_2({\bf k})$, and so on.
We remark that the expression of the probability $P_{in}({\bf k'}|{\bf k})$ is not
a necessary ingredient of the calculation. In fact, the sum on the right-hand side of 
Eq.~\ref{pagerank2} is just the average value of  
$\bar{p}_{n-1}({\bf k^\prime})/{k^\prime}_{out}$
among all predecessors of vertices with degree ${\bf k}$. 
The algorithm leads to convergence  
within a few iterations (we never needed more than $20$). 
In Fig.~\ref{fig2} we compare  
the values of $\bar{p}({\bf k})$ calculated from Eq.~\ref{pagerank2} 
with the corresponding empirical values. Here we averaged 
$\bar{p}({\bf k})$ over out-degree,  
so it only depends on the in-degree $k_{in}$. 
The variation of $\bar{p}({\bf k})$ with $k_{out}$ (for fixed $k_{in}$) 
turns out to be very small. 
The scatter plots of Fig.~\ref{fig2} show that the 
mean field approximation gives excellent results: 
the points are very tightly concentrated  
about each frame bisector, drawn as a guide to the eye.

\begin{figure}[t]
\centering
\includegraphics[width=\columnwidth]{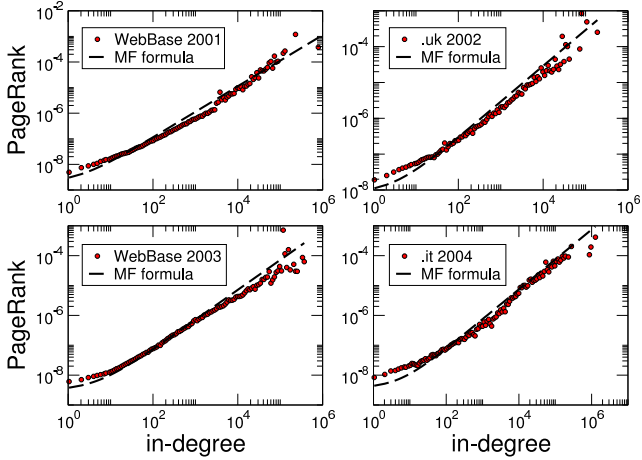} 
\caption{\label{fig3}PageRank versus in-degree; the dashed line is the approximation given by
the closed formula of Eq.~\ref{cleq}.}
\end{figure}

We now analyze explicitly the relation between PageRank and in-degree. To plot
the function $\bar{p}({k_{in}})$ directly is not very helpful because the 
wide fluctuations of PageRank within each degree class would mistify the
pattern for large values of $k_{in}$.
The best thing to do is to 
average PageRank within bins of in-degree. As both PageRank and in-degree 
are power-law distributed, we decided to use logarithmic bins; 
the multiplicative factor for the bin size is $1.3$. 
The resulting patterns for our four Web samples are presented in Fig.~\ref{fig3}.
The empirical curves are rather smooth, and show that the average PageRank 
(per degree class) is an increasing function of in-degree. The relation between the
two variables is approximately linear for large in-degrees. This is exactly 
what we would expect if the degrees of pages were uncorrelated with those 
of their neighbors in the Web graph. In such a case the relation 
between PageRank and in-degree is given by Eq.~\ref{cleq}. Indeed,
the comparison of the empirical data with the curves of Eq.~\ref{cleq} in Fig.~\ref{fig3} 
is quite good for all data sets. We infer that the Web graph is an essentially 
uncorrelated graph; this is confirmed by direct measurements of degree-degree 
correlations in our four Web samples~\cite{SantoWWW06decoding}. 
What is most important, the average PageRank of
a page with in-degree $k_{in}$ is well approximated by the simple expression of 
Eq.~\ref{cleq}. The possible applications of this result are examined in the next 
section.

\begin{figure}
\centering
\includegraphics[width=\columnwidth]{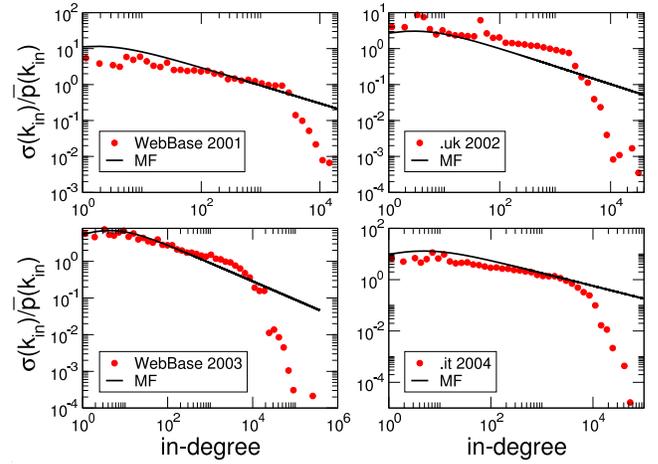} 
\caption{\label{fig4} Coefficient of variation of PageRank versus in-degree.}
\end{figure}

Let us analyze the empirical fluctuations of PageRank about its 
mean value. We anticipated in Section~\ref{fluct} that we expect large  
fluctuations for small values of $k_{in}$, due to the large value of the 
second momentum of the in-degree distribution, and that the relative size 
of the fluctuations should decrease as $k_{in}$ increases (Eq.~\ref{fleq}). 
Fig.~\ref{fig4} confirms our prediction. We plotted the coefficient of variation 
${\sigma({k_{in}})}/\overline{p}({k_{in}})$ as a function of $k_{in}$,  
once again averaging over out-degree. The trend is clearly decreasing 
as $k_{in}$ increases. The fluctuations of the data points are due to 
degree-degree correlations 
(even if they are small, they are not completely negligible). We also derived 
mean field estimates for the coefficient of variation. Rather than solving the 
complete Eq.~\ref{compfleq}, we used the coefficient of variation 
for an uncorrelated network, given by the ratio between $\sigma({k_{in}})$ from 
Eq.~\ref{simpfl} and $\overline{p}({k_{in}})$ from Eq.~\ref{cleq}. 
Nevertheless, the agreement between our approximated estimates and the empirical 
results in Fig.~\ref{fig4} is very good except for high $k_{in}$,  
where we have an insufficient number of points in each degree class leading 
to high fluctuations. 

\begin{figure}
\centering
\includegraphics[width=\columnwidth]{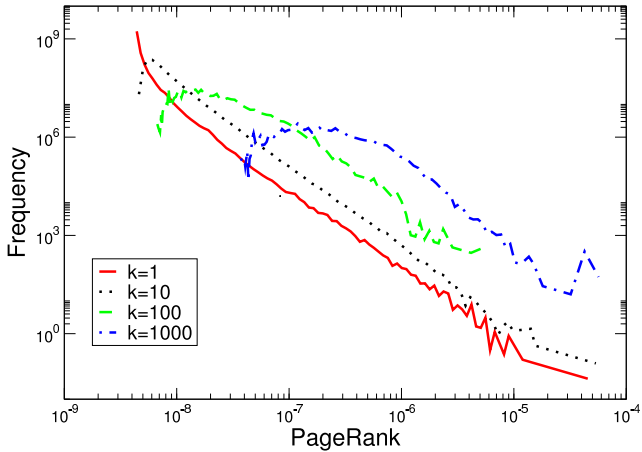} 
\caption{\label{fig5} Distributions of PageRank for four degree 
classes in the link graph from the WebBase 2003 crawl.}
\end{figure}

Finally, we test our prediction for the distribution of PageRank within a degree class. 
In Fig.~\ref{fig5} we plot the PageRank distributions for four classes, corresponding to in-degree
$1$, $10$, $100$ and $1000$. The data refer to the WebBase sample of 2003, but we found the same 
trend for the other three data sets. 
We see that for low in-degrees ($k_{in}=1, 10$) the distribution is a power law.
The exponent is essentially the same ($2.3\pm 0.1$, as for the other samples). 
However, for higher degrees ($k_{in}=100, 1000$), 
the distribution changes from a power law to a hybrid distribution between a Gaussian
and a power law. The Gaussian is signaled by the peak, which in the double logarithmic 
scale of the plot appears quite flat; the power law is manifest in the long tail 
of the distribution. The power law tails are all approximately parallel to each other, 
i.e., the exponent is the same for all curves.

\section{Applications to the live Web}
\label{apps}

We have seen that the average PageRank of a page with in-degree $k_{in}$ 
can be well approximated by the closed formula in Eq.~\ref{cleq}. We have also found 
that PageRank fluctuations about the average become less important for larger $k_{in}$. 
These two results suggest that for large enough $k_{in}$ the 
PageRank of a page with $k_{in}$ in-links only depends on $k_{in}$, and that
Eq.~\ref{cleq} should give at least the correct order of magnitude for its PageRank.
To use Eq.~\ref{cleq} for the Web we need to know the total number $N$ 
of Web pages indexed by Google and their 
average degree $\langle k_{in}\rangle$. The size of the Google index was 
published until recently. We use the last reported number, $N \simeq 8.1\times 10^9$.
The average degree is not known; the best we can do is extract it from 
samples of the Web graph. Our data sets do not deliver a unique value for 
$\langle k_{in}\rangle$, but they agree on the order of magnitude (see Table~\ref{tab1}). 
Hereafter we use $\langle k_{in}\rangle = 10$.

\begin{figure}
\centering
\includegraphics[width=\columnwidth]{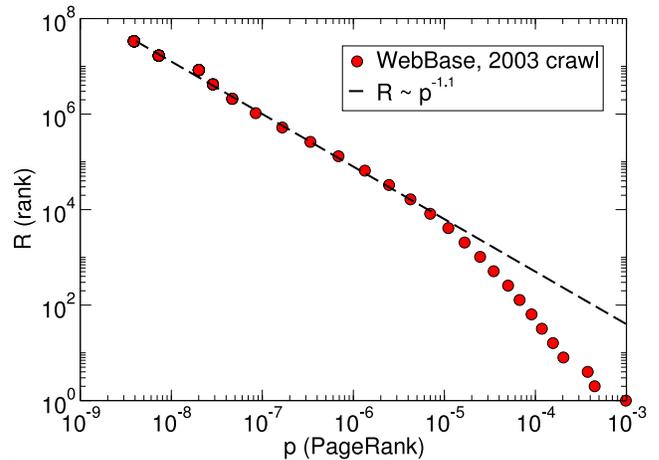} 
\caption{\label{fig6} Dependence of the rank of a page on its PageRank value.}
\end{figure}

In this section we want to see whether Eq.~\ref{cleq} can be useful in the live Web.
Ideally we should compare the PageRank values of a list of Web pages with the corresponding 
values derived through our formula. Unfortunately the real PageRank values 
calculated by Google are not accessible, so we need a different strategy. 
The simplest choice is to focus on rank rather than PageRank.
We know that Google ranks Web pages according to their PageRank values as well as other
features which do not depend on Web topology. The latter features are not disclosed; 
in the following we disregard them and assume for simplicity that the Google 
ranking of a Web page exclusively depends on its PageRank value. 
There is a simple relation between the PageRank $p$ of a Web page and 
the rank $R$ of that page. The Zipf function $R(p)$ is simply proportional to the
cumulative distribution of PageRank. 
Since the PageRank distribution is approximately a power law with 
exponent $\beta \simeq 2.1$ (see Section~\ref{results}), we find that
\begin{equation}
R(p) \simeq A \, p^{-\alpha},
\label{rR}
\end{equation}
where $\alpha = \beta - 1 \simeq 1.1$ 
and $A$ is a proportionality constant. 
Eq.~\ref{rR} can be empirically tested. Fig.~\ref{fig6} shows the 
pattern for the WebBase sample of 2003. The ansatz of Eq.~\ref{rR} (with $\alpha = 1.1$) 
reproduces the data for over three orders of magnitude. 
The rank $R$ referred to above is the global rank of a page
of PageRank $p$, i.e., its position in the list containing  
all pages of the Web in decreasing order of PageRank.
More interesting for information providers and search engine marketers 
is the rank within hit lists returned for actual queries, where only a 
limited number of result pages appear. We need a criterion to pass 
from the global rank $R$ to the rank $r$ within a query's hit list. 
A page with global rank $R$ could appear at any position $r=1,2,\dots,n$ 
in a list with $n$ hits. 
In our framework pages differ only by their PageRank values (or, equivalently, 
by their in-degrees), as we neglect all semantic features. Therefore we can assume
that each Web page has the same probability to appear in a hit list.  
This is a strong assumption, but even if it may fail to describe what happens 
at the level of an individual query, it is a fair approximation when one 
considers a large number of queries. 
Under this hypothesis the probability distribution of the possible positions 
is a Poissonian, and the expected local rank $r$ of a page with global rank $R$ 
is given by the mean value:
\begin{equation}
r = R \frac{n}{N}.
\label{rloc}
\end{equation}

Now it is possible to test the applicability of Eq.~\ref{cleq} to the Web.
We are able to estimate the rank of a Web page within a hit list if we know 
the number of in-links $k_{in}$ of the page and the 
number $n$ of hits in the list. The procedure consists of three simple steps:
\newpage
\begin{enumerate}
\item from $k_{in}$ we calculate the PageRank $p$ of the page according to Eq.~\ref{cleq};
\item from $p$ we determine the global rank $R$ according to Eq.~\ref{rR};
\item from $R$ and $n$ we derive the local rank $r$ according to Eq.~\ref{rloc}.
\end{enumerate}

\begin{figure}[t]
\centering
\includegraphics[width=\columnwidth]{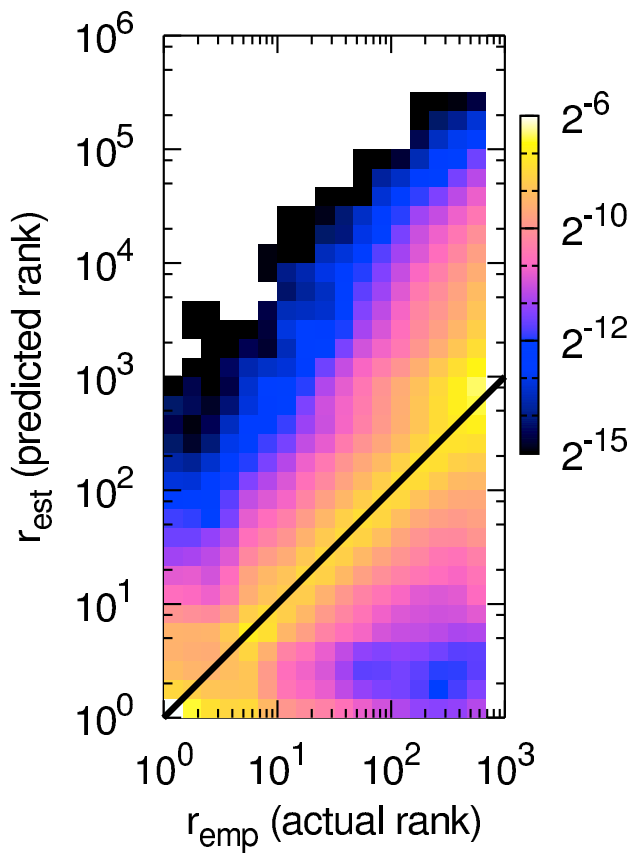} 
\caption{\label{fig7} Density map of the scatter plot between predicted rank 
$r_{est}$ and actual rank $r_{emp}$ for 65,207 queries. 
The fraction of points in each log-size bin is expressed by the color, 
also on a logarithmic scale. The diagonal guide to the 
eye corresponds to $r_{est} = r_{emp}$.}
\end{figure}

The combination of the three steps leads to the following expression of
the local rank $r$ as a function of $k_{in}$ and $n$:
\begin{equation}
r=\frac{A\,n}{(\frac{q}{N}+\frac{1-q}{N\langle k_{in}\rangle}k_{in})^{1.1}N}.
\label{rth}
\end{equation}
%
The natural way to derive the parameter $A$ would be to perform a fit of the 
empirical relation between global rank and PageRank, as we did in Fig.~\ref{fig6}. 
The result should then be extrapolated to the full Web graph. As it turns out, 
the $A$ value derived in this way strongly depends on the sample 
of the Web, so that one could do no better than estimating the
order of magnitude of $A$. On the other hand $A$ is a simple multiplicative 
constant, and its value has no effect on the dependence of the local rank $r$ 
on the variables $k_{in}$ and $n$. Therefore we decided to consider it as a free 
parameter, whose value is to be determined by the comparison with empirical data.

For our analysis we used a set of $65,207$ actual queries 
from a September 2001 AltaVista log~\cite{AltaVista}. 
We submitted each query to Google, and picked at random one of the pages of the 
corresponding hit list. For each selected page, we
stored its actual rank $r_{emp}$ within the hit list, as well
as its number $k_{in}$ of in-links, which was again 
determined through Google. The number $n$ of hits of the list was also stored.
Google (like other search engines) never displays 
more than $1000$ results per query, so we always have $r_{emp} \leq 1000$. 
From $k_{in}$ and $n$ we estimated the theoretical rank $r_{est}$ 
by means of Eq.~\ref{rth}, and compared it with its empirical counterpart $r_{emp}$. 
The comparison can be seen in the scatter plot of Fig.~\ref{fig7}. 
Given the large number of queries and the broad range of rank values,  
we visualize the density of points in logarithmic bins. 
The region with highest density is a stripe centered on the diagonal 
line $r_{est}=r_{emp}$ by a suitable choice of $A$ ($A=1.5\times 10^{-4}$). 
We conclude that the rank derived through Eq.~\ref{rth} is in most cases close 
to the empirical one. We stress that this result is not trivial, because (i) 
Web pages are not ranked exclusively according to PageRank; (ii) we are 
neglecting PageRank fluctuations; and (iii) all pages do not have the same 
probability of being relevant with respect to a query.

\section{Discussion}

The present study motivates further enquires. The mean field approach 
provides a simple functional relationship between average PageRank, 
in-degree, and degree-degree correlations.
The price one pays by using such a simple approximation is the neglect 
of the significant fluctuations of PageRank values 
around the mean field average within a degree class. 
For the majority of pages, having moderate PageRank, 
fluctuations are more important; the in-degree being equal, 
they make the difference between being linked by ``good'' 
or ``bad'' pages. A venue we intend to pursue
is to understand what makes the difference between two pages with 
the same in-degree and a very different value of PageRank, 
and how pages with higher PageRank are differently positioned 
in the complex architecture of the Web graph. 

The approach described here lends itself naturally to applications 
other than the Web, e.g., bibliometry. Commonly the quality of papers is assessed via 
the number of citations they receive, and it would be useful to be able to rank papers
with the same number of citations through their PageRank values. 
A characterization of papers leading to high PageRank fluctuations 
would be useful in this domain as well. 

A promising way to study fluctuations at a moderate price in increased complexity 
could be to use the definition of Eq.~\ref{eq1} where the value of PageRank on the 
right-hand side is substituted by the mean field approximation. 
Further work is needed in this direction.

In this paper we have quantitatively explored two key assumptions around 
the current search status quo, namely that PageRank is very different from 
in-degree due to its global nature and that PageRank cannot be easily 
guessed or approximated without global knowledge of the Web graph. We have shown 
that due to the weak degree-degree correlations in the Web link graph, PageRank 
is strongly correlated with in-degree and thus the two measures provide very 
similar information --- the PageRank factor used by search engines to rank 
pages can be effectively replaced by in-degree, especially for the most popular pages.
Further, we have introduced a general mean field approximation 
of PageRank that, in the specific case of the Web, allows to estimate PageRank from 
only local knowledge of in-degree. We have further quantified the fluctuations of 
this approximation, gauging the reliability of the estimate. Finally we have validated 
the approach with a simple procedure that predicts how actual Web pages are ranked 
by Google in response to actual queries, using only knowledge about in-degree and the 
number of query results. 



Our method has immediate application for information providers. 
For instance, the association between rank and in-degree allows one to deduce how many
in-links would be needed for a new page to achieve a given rank among all pages that deal 
with the same topic. This is an issue of crucial economic impact: 
all companies that advertise their products and services online wish for  
their homepages to belong among the top-ranked sites in their business sector. 
Suppose that someone wants their homepage $H$ to appear among the top $n$ pages 
about topic $T$. Our recipe is extremely simple and cheap, requiring 
the submission of two queries to the search engine: 
\begin{enumerate}
\item submit a query to Google (or another search engine) about topic $T$;
\item find the number $k_n$ of in-links for the $n$-th page in the resulting hit list;
\item $H$ needs at least $k_n$ in-links to appear among the first $n$ hits for topic $T$.
\end{enumerate}
Of course there are limits to this approach; we do not claim that the lower 
bound of the number of in-links can be taken as a safe guide. Indeed we have 
neglected important factors such as the role of page content in retrieving 
and ranking results, and the fluctuations of the mean field approximation of 
PageRank. 

Notwithstanding the above caveats, our results indicate that 
at least the order of magnitude should be a reliable reference point. 
This may be all that is necessary --- knowing the difference between 
the need for one thousand or one million links can be a crucial asset in 
planning and budgeting a marketing campaign. Is word of mouth sufficient, or 
is advertising required? Our approach provides a tool to answer this 
kind of questions. In making such a tool available to search engine marketers 
and information providers alike, we hope to create a more level playing field so 
that not only large and powerful organizations but also small communities 
with little or no marketing budget can make informed decisions about the management 
of their Web presence. 


\section{Acknowledgments}

We thank the members of the Networks and Agents Network at
IUB, especially Alessandro Vespignani and Mariangeles Serrano, 
for helpful discussions. We are grateful to Google for extensive use of 
its Web API, to the WebBase and WebGraph projects for their crawl
data, and to AltaVista for use of their query logs.  
This work is funded in part by a Volkswagen Foundation grant to SF, 
by the Spanish government's DGES grant FIS2004-05923-CO2-02 to MB, 
by NSF Career award 0348940 to FM, and by the Indiana University School 
of Informatics.


\newpage
\begin{thebibliography}{10}

\bibitem{AltaVista}
{AltaVista}.
\newblock http://www.altavista.com.

\bibitem{Amento00}
B.~Amento, L.~Terveen, and W.~Hill.
\newblock Does ``authority'' mean quality? {P}redicting expert quality ratings
  of {Web} documents.
\newblock In {\em Proc. 23rd ACM SIGIR Conf. on Research and Development in
  Information Retrieval}, pages 296--303, 2000.

\bibitem{Brin98}
S.~Brin and L.~Page.
\newblock The anatomy of a large-scale hypertextual {W}eb search engine.
\newblock {\em Computer Networks}, 30(1--7):107--117, 1998.

\bibitem{millozzi}
D.~Donato, L.~Laura, S.~Leonardi, and S.~Millozzi.
\newblock Large scale properties of the webgraph.
\newblock {\em European Physical Journal B}, 38:239--243, 2004.

\bibitem{GoogleAPI}
{Google Web API}, 2005.
\newblock {http://www.google.com/apis}.

\bibitem{Gori05cacm}
M.~Gori and I.~Witten.
\newblock The bubble of web visibility.
\newblock {\em Communications of the ACM}, 48(3):115--117, 2005.

\bibitem{iprospect}
iProspect.
\newblock White papers, tools and search engine marketing research studies.
\newblock http://www.iprospect.com/about/ searchenginemarketingwhitepapers.htm.

\bibitem{WebGraph}
{Laboratory for Web Algorithmics (LAW), University of Milan}, 2005.
\newblock {http://webgraph.dsi.unimi.it}.

\bibitem{ubicrawler}
{Laboratory for Web Algorithmics (LAW), University of Milan, and Istituto di
  Informatica e Telematica of CNR}, 2005.
\newblock {http://ubi.iit.cnr.it/projects/ubicrawler}.

\bibitem{naka}
I.~Nakamura.
\newblock Large scale properties of the webgraph.
\newblock {\em Physical Review E}, 68:045104, 2003.

\bibitem{Upfal2002}
G.~Pandurangan, P.~Raghavan, and E.~Upfal.
\newblock Using pagerank to characterize {Web} structure.
\newblock In {\em Proc. 8th Annual International Conference on Combinatorics
  and Computing (COCOON)}, pages 330--339, Singapore, 2002. Springer-Verlag.

\bibitem{Cho05WebDB}
F.~Qiu, Z.~Liu, and J.~Cho.
\newblock Analysis of user web traffic with a focus on search activities.
\newblock In {\em Proc. International Workshop on the Web and Databases
  (WebDB)}, 2005.

\bibitem{SantoWWW06decoding}
M.~Serrano, A.~Maguitman, M.~{Bogu\~{n}\'{a}}, S.~Fortunato, and A.~Vespignani.
\newblock Decoding the structure of the {WWW}: Facts versus bias.
\newblock Technical report, Indiana University School of Informatics, 2005.
\newblock Submitted to WWW2006.

\bibitem{sew-seo}
D.~Sullivan.
\newblock Intro to search engine optimization.
\newblock http://searchenginewatch.com/webmasters/ article.php/2167921.

\bibitem{NielsenAug05}
D.~Sullivan.
\newblock Nielsen//netratings search engine ratings.
\newblock http://searchenginewatch.com/reports/article.php/ 2156451, August
  2005.

\bibitem{WebBase}
{WebBase Project}, 2005.
\newblock {http://www-diglib.stanford.edu/\~{}testbed/doc2/WebBase/}.

\bibitem{websidestory}
Websidestory, May 2005.
\newblock Cited by Search Engine Round Table,
  http://www.seroundtable.com/archives/001901.html. According to this source,
  Websidestory Vice President Jay McCarthy announced at the Search Engine
  Strategies Conference (Toronto 2005) that the number of page referrals from
  search engines has surpassed those from other pages.

\end{thebibliography}
\end{document}